\documentstyle[aps,prb,multicol,epsf]{revtex}

\begin{document}

\title{Far-infrared magnetotransmission of
YBa$_2$(Zn$_x$Cu$_{1-x}$)$_3$O$_{7-\delta}$
thin films}

\author
{Jan Kol\'a\v{c}ek, Roman Tesa\v{r}, Zden\v{e}k \v{S}im\v{s}a}

\address{Institute of Physics ASCR, 
  Cukrovarnick\'a 10, 16253 Prague 6, 
  Czech Republic}
\maketitle
\begin{abstract}
Measurements of the far infrared magnetotransmission of
YBa$_2$(Zn$_x$Cu$_{1-x}$)$_3$O$_{7-\delta}$
thin film ($x \sim$ 0.025) deposited on a wedged MgO substrate are reported.
The  application of magnetic field perpendicular to the
$ab$ plane produces at low temperature a linear increase of transmission
for frequencies below $\sim$ 30 cm$^{-1}$.
We present a model of high frequency vortex dynamics which qualitatively
explains these results.
\end{abstract}

\begin{multicols}{2}
There is a widespread agreement that high frequency conductivity is
influenced by a vortex dynamics. It is estimated that the main
characteristic frequencies of the vortex system
are in the far infrared (FIR) region. It makes FIR
magneto-optics a suitable tool to study
vortex dynamics. In spite of its importance only a few
experimental data can be found in the literature
(see e.g. \cite 
{99Liu,98Mallozzi,96Lihn,96Wu,95Eldridge,94Choi,94Shimamotoa,94Shimamotob,94Drew}). 
Here, the results of the FIR transmission measurements
made on a Zn doped YBaCuO  thin film in magnetic fields up to 5.4 T
are reported and explained by a new vortex dynamics model.

\begin{figure}[t]
\epsfxsize=75mm
\epsffile{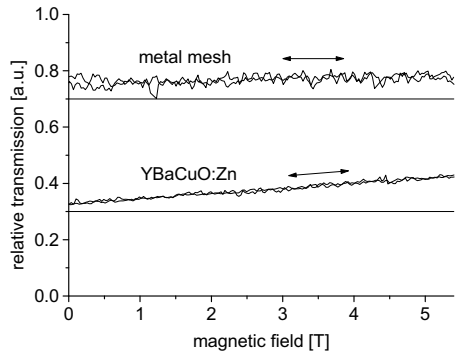}
\columnwidth 85mm
\caption{
The dependence of transmission on the external magnetic field for
YBaCuO:Zn sample and the metal mesh measured
at temperature of 3K and frequency of 13 $\mathrm{cm}^{-1}$.}
\label{fig1}
\end{figure}

The measurements were performed at several frequencies
from 13.5 to 142.9 $\mathrm{cm}^{-1}$  on the high quality thin film 
YBa$_2$(Zn$_x$Cu$_{1-x}$)$_3$O$_{7-\delta}$
prepared by the laser ablation deposition on a wedged MgO substrate.
From the critical temperature $T_{\mathrm c} = 64.3 \mathrm K$
the parameter $x$ was estimated as $x \sim 0.025$.
The dependence of the far-infrared transmission on the
magnetic field apllied perpendicularly to the $ab$ plane
was acquired on our  laser based spectrophotometer FIRM
described in details earlier \cite{93Kolacek}.
The radiation transmitted through the sample kept at temperature 3 K
is measured by a silicon bolometer; stability of the laser line
is monitored by a pyroelectric detector.
Measurement of a wire mesh
proved that bolometer is not influenced by a stray magnetic field.
The field was swept slowly at a rate of 1 T/min.
Equilibrium conditions were preserved as evidenced by
the coincidence
of the relative transmittance monitored with increasing and decreasing 
magnetic field - see Fig.\ \ref{fig1}.

\begin{figure*}[t]
\epsfxsize=75mm
\epsffile{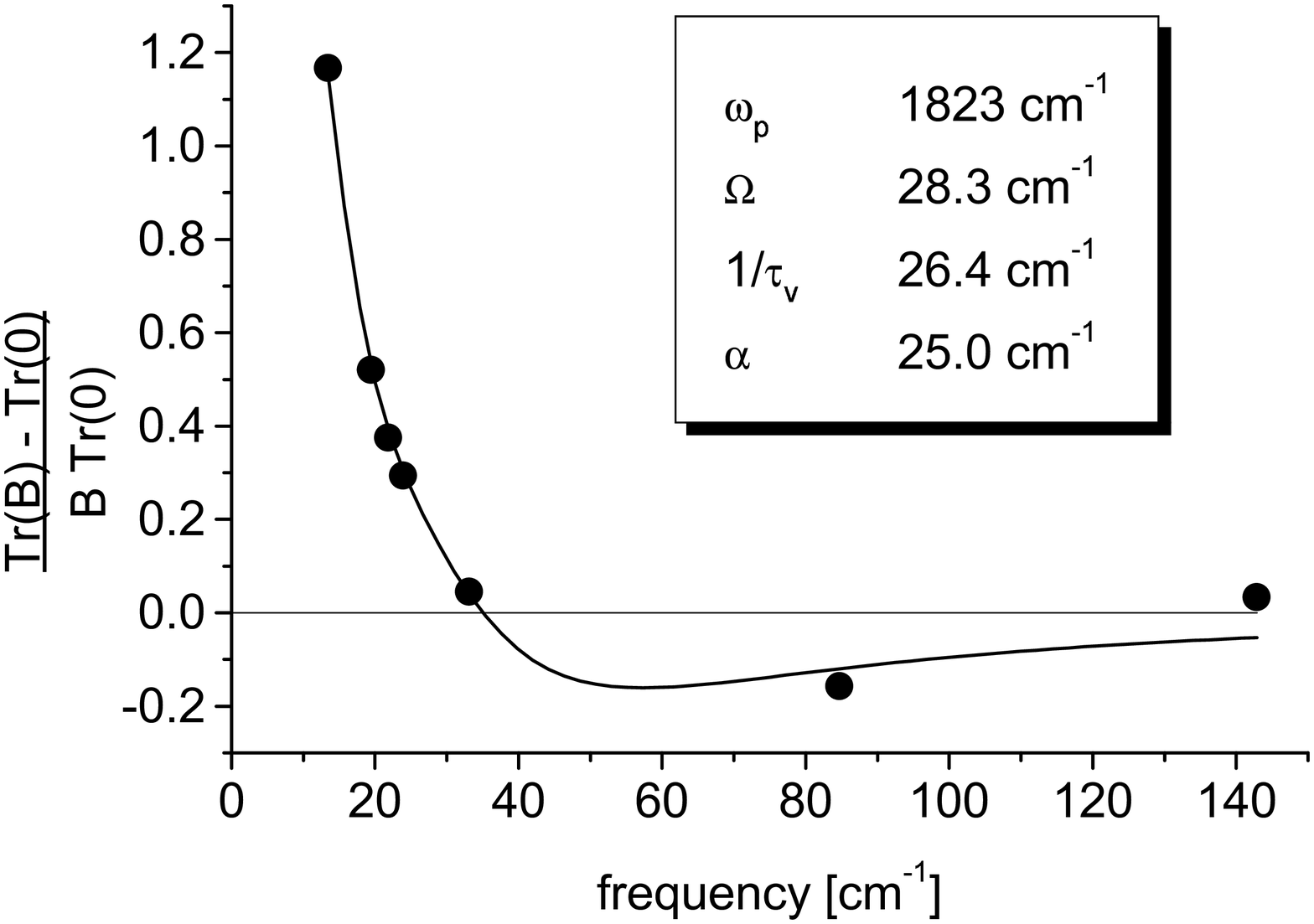}
\columnwidth 85mm
\caption{The frequency dependence of the measured
(points) and calculated (solid line) transmission slopes.}
\label{fig2}
\end{figure*}

The frequency dependence of the transmission slope
${Tr(B)-Tr(0)} \over {B Tr(0)}$ is displayed in Fig.\ \ref{fig2}.
For frequencies below $30 \mathrm{cm}^{-1}$
it is positive, at frequency of $ 85\mathrm{cm}^{-1}$
it is negative, whereas at $143 \mathrm{cm}^{-1}$
hardly any field dependence was observed.

This behaviour can qualitatively be explained
by our model of vortex dynamics \cite{99Kolacek}.
At low temperatures the influence of normal state fluid
may be neglected, so that the electrodynamic responce is determined
by vortices and superconducting fluid.
Equations of motion of these two mutually interacting subsytems
may be written as

\begin{equation} 
m {\dot {\bf v}}_{\mathrm s} = e{\bf E}
     - \omega_{\mathrm c} ({\bf v}_{\mathrm s} - {\bf v}_{\mathrm L})
     \times {\bf z}
\label{eqmotions}
,\end{equation}
\begin{equation}
m {\dot {\bf v}}_{\mathrm L} = - \alpha ^2 {\bf r}_{\mathrm L}
     - \frac {1} {\tau_{\mathrm L}}  {\bf v}_{\mathrm L}
     + \Omega ({\bf v}_{\mathrm s} - {\bf v}_{\mathrm L}) \times {\bf z}
\label{eqmotionv}
, \end{equation}
where ${\bf r}_{\mathrm L}$ and ${\bf v}_{\mathrm L}$ are the
position and velocity of the vortex lattice, while
${\bf v}_{\mathrm s}$ is velocity of the superconducting fluid.
The interaction of superconducting fluid and the vortex lattice is
mediated by the Magnus force with $\Omega$ and $\omega_{\mathrm c}$
being angular frequency of the circular vortex motion and
cyclotron frequency of the superconducting fluid.
Pinning force and vortex damping are characterised
by pinning frequency $\alpha$
and vortex relaxation time $\tau_{\mathrm v}$.

Solving these two equations for the right (+) and left (-) hand polarized mode
and using the expression for the current
${\bf j} = n_{\mathrm s}e {\bf v}_{\mathrm s}  = \sigma \bf E$
the conductivity $\sigma^{\pm}$ may be expressed as
\begin{equation}
\sigma ^{\pm} =
    \epsilon_0 \omega_p^2
    \frac
    {\pm\omega / \tau_v-{\mathrm i}(\alpha^2-\omega^2 \pm\Omega\omega)}
    {(\pm\omega-\omega_{\mathrm c})
    (\alpha^2-\omega^2 \pm{\mathrm i}\omega/\tau_{\mathrm v}) + \Omega\omega^2}
\label{conductivity}
\end{equation}
where $\omega_{\mathrm p}$ is plasma frequency.
For linearly polarized light used in the experiment the transmission
coefficient may be expressed as $Tr = |\frac{1}{2} (t^+ + t^-)|^2$
where $t^\pm$ are the transmission amplitudes given by
$t^\pm=\frac {n+1}
       {Z_0 d \sigma^\pm +n+1}$.
Here $d = 150$ nm is the film thickness, $n =4 $ is refractive index of the
substrate and $Z_0 =377 \mathrm{ohm}$ is impedance of the free space.
Using parameters displayed as insert in Fig. 2,
the experimental and calculated data are
in good agreement.


\acknowledgements
The authors thank to M.Jelinek for the sample preparation.
This work was supported by M\v{S}MT program "KONTAKT ME 160 (1998)".


\end{multicols}
\end{document}